# Room-temperature tunable tunneling magnetoresistance in $Fe_3GaTe_2/WSe_2/Fe_3GaTe_2$ van der Waals heterostructures

Haiyang Pan[1,2], Anil Kumar Singh[3], Chusheng Zhang[1], Xueqi Hu[1], Jiayu Shi[4], Liheng An[1], Naizhou Wang[1], Ruihuan Duan[4], Zheng Liu[4], Stuart S. P. Parkin[5], Pritam Deb[3], Weibo Gao[1,6,7,*]

[1]Division of Physics and Applied Physics, School of Physical and Mathematical Sciences, Nanyang Technological University, Singapore 637371, Singapore

[2]School of Materials Science and Engineering, Yancheng Institute of Technology, Yancheng 224051, China

[3]Department of Physics, Tezpur University (Central University), Tezpur 784028, India

[4]School of Materials Science and Engineering, Nanyang Technological University, Singapore, Singapore

[5]Max Planck Institute of Microstructure Physics, Weinberg 2, 06120, Halle (Saale), Germany

[6]The Photonics Institute and Centre for Disruptive Photonic Technologies, Nanyang Technological University, Singapore 637371, Singapore

[7]Center for Quantum Technologies, National University of Singapore, Singapore

[*]Corresponding author.

Email: wbgao@ntu.edu.sg

## Abstract

The exceptional properties of two-dimensional (2D) magnet materials present a novel approach to fabricate functional magnetic tunnel junctions (MTJ) by constructing full van der Waals (vdW) heterostructures with atomically sharp and clean interfaces. The exploration of vdW MTJ devices with high working temperature and adjustable functionalities holds great potential for advancing the application of 2D materials in magnetic sensing and data storage. Here, we report the observation of highly tunable room-temperature tunneling magnetoresistance through electronic means in a full vdW $Fe_3GaTe_2$/$WSe_2$/$Fe_3GaTe_2$ MTJ. The spin valve effect of the MTJ can be detected even with the current below 1 nA, both at low and room temperatures, yielding a tunneling magnetoresistance (TMR) of 340% at 2 K and 50% at 300 K, respectively. Importantly, the magnitude and sign of TMR can be modulated by a DC bias current, even at room temperature, a capability that was previously unrealized in full vdW MTJs. This tunable TMR arises from the contribution of energy-dependent localized spin states in the metallic ferromagnet $Fe_3GaTe_2$ during tunnel transport when a finite electrical bias is applied. Our work offers a new perspective for designing and exploring room-temperature tunable spintronic devices based on vdW magnet heterostructures.

# Introduction

The magnetic tunnel junction (MTJ) is the leading storage component of non-volatile magnetic random access memory (MRAM) technologies.[1-2] It consists of a thin tunnel barrier layer sandwiched between two magnetic layers, offering fast switching speed, high endurance, and low power consumption.[3] As big data and the Internet of Things continue to grow, optimizing the operation of MTJ to achieve lower energy consumption for high-density memory and faster data processing becomes crucial.[4] One effective and easily accessible approach to manipulating the MTJ is using the electric field, which is realized in ferromagnetic/ferroelectric multiferroic heterostructures.[5] The behavior and performance of the MTJ spintronic devices are significantly influenced by the interface between heterostructures.[4] Thus, achieving a high-quality interface of MTJ is of utmost importance to fully exploit its capabilities and enhance data processing speeds.

The emergence of two-dimensional (2D) van der Waals (vdW) magnets opens a promising avenue to construct vdW heterostructures with atomically sharp interface with minimal interlayer coupling,[6-14] which makes it possible to explore novel electronic control of MTJ spintronic devices.[4, 15] In recent years, notable advancements have been made in spin-valve devices based on ferromagnetic metal in full vdW MTJs with tunnel barrier hBN, $MoS_2$, and InSe, etc.[16-21] Recent studies have reported tunable TMR through electronic means in vdW heterostructures at low temperature.[16] However, achieving electrical control of TMR operation at room temperature remains an ongoing challenge, and the room-temperature tunable TMR has not been achieved in vdW heterostructures until now. Nevertheless, the discovery of 2D vdW ferromagnetic (FM) metal Fe$_3$GaTe$_2$,[22] which exhibits strong ferromagnetism above room temperature (Curie temperature ≈ 350-380 K) and robust large perpendicular magnetic anisotropy, opens up the possibility of room temperature spin manipulation in vdW spin devices.[23-25]

In this study, we present the observation of the tunable TMR at room temperature in vertical vdW MTJ utilizing Fe$_3$GaTe$_2$/WSe$_2$/Fe$_3$GaTe$_2$ (FGT/WSe$_2$/FGT) heterostructures. The device configuration consists of Fe$_3$GaTe$_2$ layers as both the top and bottom ferromagnetic electrodes, while few-layer WSe$_2$ serves as the tunnel barrier. Through the application of an DC electric bias, the TMR can be tunable from positive to negative and then back to positive at room temperature. The tunable TMR can be attributed to the energy-dependent nature of the spin polarization in Fe$_3$GaTe$_2$, and the high-energy localized spin states of the metallic ferromagnet Fe$_3$GaTe$_2$ with different polarized spin sources can be accessed and participate in the tunnel process by applying a DC bias. Furthermore, the TMR signal can be reliably detected even with a current as low as 1 nA at room temperature. Remarkably, the MTJ device exhibits a significant TMR of 340% at low temperatures with the substantial high spin polarization up to 80%.

**Results and discussion**

The optical microscopy image of FGT/WSe$_2$/FGT MTJ is displayed in Figure S1, where the few-layer WSe$_2$ is sandwiched between two ferromagnetic Fe$_3$GaTe$_2$ electrodes. To enable distinct TMR states during a field sweep, different layer thicknesses of Fe$_3$GaTe$_2$ are utilized for the top and bottom ferromagnetic electrodes. This variation in thickness results in different coercive fields, ensuring the emergence of diverse TMR states as the magnetic field is swept. Atomic force microscopy (AFM) measurements indicate that the thickness of the bottom Fe$_3$GaTe$_2$ layer, WSe$_2$ tunnel barrier, and top Fe$_3$GaTe$_2$ layer is about 18 nm, 7 nm, and 40 nm, respectively. The magnetic field is applied vertically, aligned with the easy axis of Fe$_3$GaTe$_2$. The tunnel resistance is measured by applying a constant AC current through the top and bottom Fe$_3$GaTe$_2$ electrodes, and measuring the voltage drop at the same time. The temperature-dependent tunnel resistance, DC *I-V* curve, and spin-valve magnetoresistance behavior of the FGT/WSe$_2$/FGT MTJ are also presented in Figure S1. The nonlinear *I-V* curve confirms the excellent tunneling characteristics of the MTJ, which is different from other reported linear *I-V* curves of Fe$_3$GaTe$_2$ MTJs with MoSe$_2$ and MoS$_2$ tunnel barriers.[26-27] The perfect tunnel barrier and large TMR of 50% at room temperature indicate the high-quality interface of our fabricated full vdW MTJ.

As the perfect TMR has been established in vdW FGT/WSe$_2$/FGT MTJ, we then tried to tune the TMR at room-temperature through applying a DC bias across the MTJ. The room-temperature magnetoresistance (MR) curves at various positive and negative current bias are displayed in Figure S2 and Figure S3. The spin valve behavior of sudden resistance switching can be observed, when a magnetic field is swept in the positive or negative direction. This phenomenon arises from the sequential switching of the magnetization direction in the top and bottom Fe$_3$GaTe$_2$ ferromagnets, resulting in the observed resistance transition. For the MR without DC bias current (0 µA), the low resistance state corresponds to the parallel alignment of the two magnetization directions, while the high resistance state corresponds to the antiparallel alignment. To assess the TMR quantitatively, we define TMR=$(R_{AP}-R_P)/R_P$, where $R_{AP}$ and $R_P$ represent the resistances in the antiparallel and parallel magnetization alignments, respectively. Figure 1A and 1B show the room-temperature TMR curves at various positive and negative current bias, respectively. At the DC bias current of 0 µA, the TMR exceeds 50%, surpassing the performance of most MTJ devices based on $Fe_3GaTe_2$.[26-28] When gradually increase the amplitude of the DC current bias from 0, the positive TMR decreases, eventually disappearing as the DC current bias reaches +/-3 µA. Further increasing DC bias amplitude, the TMR evolves into negative regime and the magnitude of the negative TMR reaches maximum at +/-15 µA. Subsequently, the positive TMR reappears when the current bias offset surpasses -30 µA. Thus, the current bias can effectively modulate the magnitude and sign of TMR at room-temperature. The tuning effect of DC current bias on TMR is more prominent at 2 K as shown in Figure S4 and Figure S5. The TMR versus DC current bias at 300 K and 2 K are displayed in Figure 1C. Since the DC current bias applied to the MTJ produces an effective DC bias voltage, the relationship between bias voltage and current can be obtained from the *I-V* curve (Figure S1C). Figure S6 displays the TMR versus DC voltage bias at 300 K and 2 K. The evolution of TMR at positive and negative bias voltage is almost the same, indicating the symmetry interface of FGT/WSe$_2$/FGT MTJ. Notably, the TMR without DC current bias at 2 K reaches an exceptional value of 340%, which surpasses that of the recently reported vdW MTJ devices based on $Fe_3GaTe_2$[23, 26-28] and $Fe_3GeTe_2$[16-21] at low temperature.

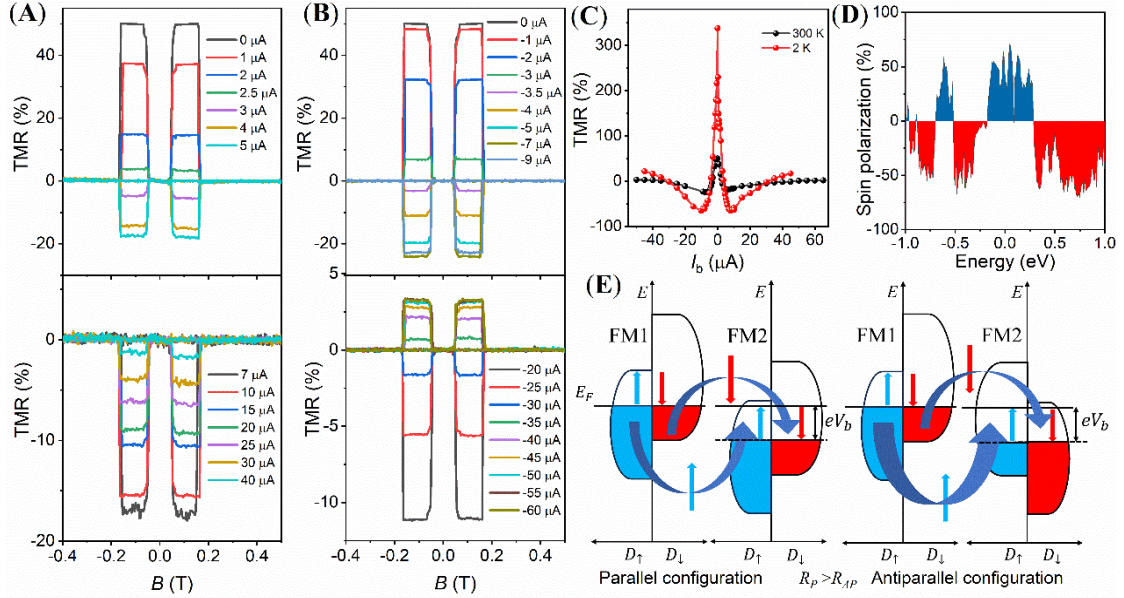

Figure 1. DC bias tunable room-temperature TMR of FGT/WSe$_2$/FGT MTJ. (A), (B), TMR measured at the various positive (A) and negative (B) DC current bias ($I_b$) at 300 K. (C) TMR versus DC current bias at 2 K and 300 K. The data point is extracted from the TMR measurement of different DC current bias at 300 K and 2 K. (D) Energy-dependent spin polarization percentage of Fe$_3$GaTe$_2$ obtained from band calculations. The spin polarization is substantially sensitive to energy and switches its polarity several times from positive to negative and vice-versa at high energies (below Fermi energy). (E) Energy-band alignments of FGT/WSe$_2$/FGT MTJ at non-zero DC bias voltage ($V_b$) in the parallel and antiparallel of two Fe$_3$GaTe$_2$ magnetization directions.

The tunable TMR can be attributed to the band alignment of two Fe$_3$GaTe$_2$ FM electrodes under DC bias, which enables the high-energy localized spin states to contribute to the tunneling process. The calculation of the electronic structure of Fe$_3$GaTe$_2$ was performed. The density of states (DOS) of spin up ($D_\uparrow$) and spin down ($D_\downarrow$) change greatly below $E_F$ (Figure S7). Figure 1D shows the energy-dependent spin polarization of Fe$_3$GaTe$_2$ estimated from the spin-polarized DOS. The spin polarization of Fe$_3$GaTe$_2$ at high energy (below $E_F$) can be substantially modified by the localized spin states. The spin polarization is positive near $E_F$, but became negative in the range of -0.2 eV to -0.5 eV, and then positive again. It's worth noting that few-layer WSe$_2$ serves solely as a tunnel barrier despite its layer-dependent band structure. The layer of WSe$_2$ has negligible impact on the spin-polarized electronic structure of Fe$_3$GaTe$_2$ layers. For the MTJ, the conductance $G$ can be described by $G \propto D_\uparrow^1 D_\uparrow^2 + D_\downarrow^1 D_\downarrow^2$, where $D_\uparrow^1$, $D_\downarrow^1$ and $D_\uparrow^2$, $D_\downarrow^2$ represent the spin up and spin down state for ferromagnetic electrodes 1 (FM1) and 2 (FM2).[16, 29-30] The TMR value arises from the disparity in conductance between the parallel and antiparallel configurations of FM1 and FM2. Figure 1E illustrates the schematic diagrams depicting electron tunnel transport in the FGT/WSe$_2$/FGT MTJ. In the absence of a bias voltage, only the itinerant spin states at $E_F$ contribute to the tunneling process, and the DOS of electronic states in both FM1 and FM2 is primarily influenced by the majority spins. As a result, $R_{AP}$ is higher than $R_P$, leading to positive TMR. When a finite DC bias voltage ($V_b$) is applied to the MTJ, $E_F$ of FM2 aligns with the filled states of FM1 at -e$V_b$, while $E_F$ of FM1 aligns with the empty states of FM2 at e$V_b$. In the energy range ($E_F$-e$V_b$, $E_F$), an energy window comprising electronic states of FM1 can tunnel into the available empty electronic states of FM2. This enables high-energy localized spin states to contribute to the tunnel conduction. When the DOS of filled electronic states in FM1 is dominated by the minority spins, and the DOS of empty states in FM2 is dominated by the majority spins, $R_P$ is expected to exceed $R_{AP}$, resulting in a negative TMR. Therefore, the observed reversals in TMR polarity can be attributed to the significant difference in spin polarization between the spin injection and detection layers of $Fe_3GaTe_2$.

To investigate the influence of operating current on the performance of the TMR, we conducted measurements of magnetoresistance by applying various AC currents to the MTJ. Figure 2A and 2B depict the TMR obtained with currents ranging from 0.1 nA to 5 µA at 2 K and 300 K, respectively. Notably, the measured TMR at 1 nA remains highly robust signal at both low and high temperatures. Additionally, Figure S8 illustrates the MR measured below 1 nA. The robust TMR observed at such low currents and at room temperature distinguishes this study from other reported vdW spin valve devices.[18-19, 21, 26-27] The TMR versus applied currents at 2 K and 300 K are presented in Figure 2C and 2D. It is evident that the large TMR values can be maintained over a wide range of currents. At the low temperature of 2 K, the TMR exhibits a slight decrease but it still exceeds 300% as the applied current increases from 0.1 nA to 100 nA. However, beyond 100 nA, the TMR decreases almost linearly with the current. This linear dependence can be ascribed to the scattering of higher energy electrons by localized trap states at the heterostructure interfaces.[18, 27] At room temperature, the TMR shows a negligible decrease with the current increasing, which is similar to the behavior observed at low temperatures in the low current region. Moreover, TMR values larger than 50% are maintained up to 0.5 µA. The wide range of operating current bias indicates that the Joule heating effect of the $FGT/WSe_2/FGT$ MTJ is minimal in the low bias region, enabling the operation of the MTJ device at room temperature with low current.

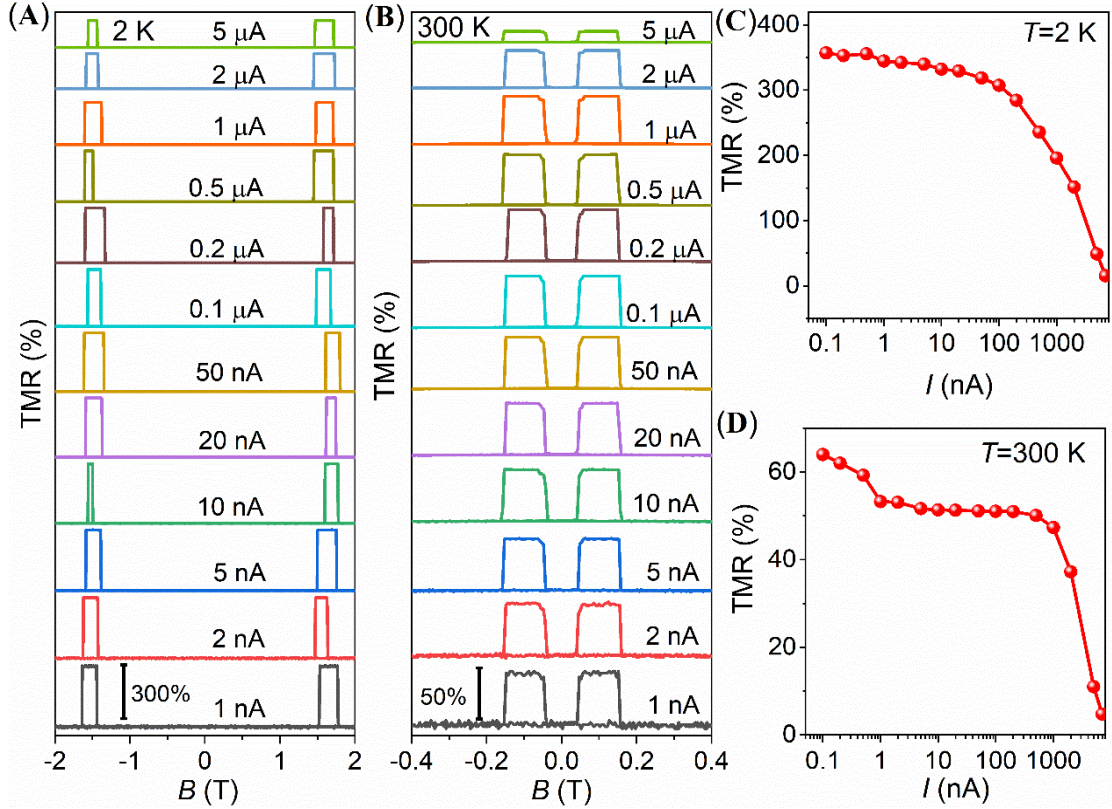

Figure 2. Measuring current dependence of TMR in the FGT/WSe$_2$/FGT MTJ. (A) The offset plot of TMR curves measured with various AC currents (*I*) ranging from 1 nA to 5 µA at 2 K. (B) The TMR curves of various AC currents at 300 K. (C) The extracted TMR ratio of 2 K as a function of measuring current *I*. (D) The extracted TMR ratio of 300 K as a function of measuring current *I*.

To gain further insight into the evolution of the TMR with temperature, we conducted TMR measurements at various temperatures. Figure 3A and 3B illustrate the TMR curves obtained, demonstrating that the prominent spin valve effect persists even at temperatures exceeding room temperature, specifically up to 340 K. This observation aligns with the Curie temperature of Fe$_3$GaTe$_2$ crystals.[22] The original MR curves are presented in Figure S9. As the temperature increases, the switching field experiences a decrease, primarily due to the reduced coercive field of the top and bottom Fe$_3$GaTe$_2$ ferromagnetic electrodes. The temperature-dependent TMR is depicted in Figure 3C. The TMR value exhibits a monotonous decrease from its highest value of 340% at 2 K to 50% at 300 K. This reduction in TMR is attributed to the decline in the spin polarization ($P$) of $Fe_3GaTe_2$ with the temperature rising. According to the Julliere model,[31] the relationship between TMR and spin polarization can be described by TMR=$2P_1P_2/(1 - P_1P_2)$, where $P_1$ and $P_2$ represent the spin polarizations of the top and bottom $Fe_3GaTe_2$ layers, respectively. Since the top and bottom layers in FGT/WSe$_2$/FGT MTJ consist of the same ferromagnetic crystals, we can assume that $P_1=P_2=P$, denoting the spin polarization. The temperature dependence of $P$ is presented in Figure 3D. At low temperatures, the spin polarization can even reach 80%, which corresponds to a majority spin percentage of 90% and a minority spin percentage of 10%. The observed high spin polarization in our MTJ device surpasses reported values for $Fe_3GaTe_2$-based spin valves.[23, 26-27] As the temperature rises, $P$ monotonically decreases from 80% at low temperatures to 50% at 300 K. The temperature dependence of spin polarization can be well described by the equation $P(T) = P_0(1 - \alpha T^{3/2})$, where $P_0$ represents the spin polarization at 0 K, and $\alpha$ is a constant used to characterize the rate of decrease of $P$ in the material with temperature.[17, 32] The fitted value of $\alpha$ is $8.1\times10^{-5}$ $K^{-3/2}$, which is approximately one order of magnitude smaller than that observed in reported spin valve devices based on $Fe_3GaTe_2$[26-28] and $Fe_3GeTe_2$[18-20]. It is noteworthy that the fitted value of $\alpha$ in our spin valve device closely resembles the values found in materials like NiFe ($3\sim5\times10^{-5}$ $K^{-3/2}$) and Co ($1\sim6\times10^{-6}$ $K^{-3/2}$).[33] This suggests that the decay of spin polarization in our FGT/WSe$_2$/FGT MTJ is significantly weaker and comparable to that of room-temperature ferromagnetic metals, owing to the high Curie temperature of $Fe_3GaTe_2$.

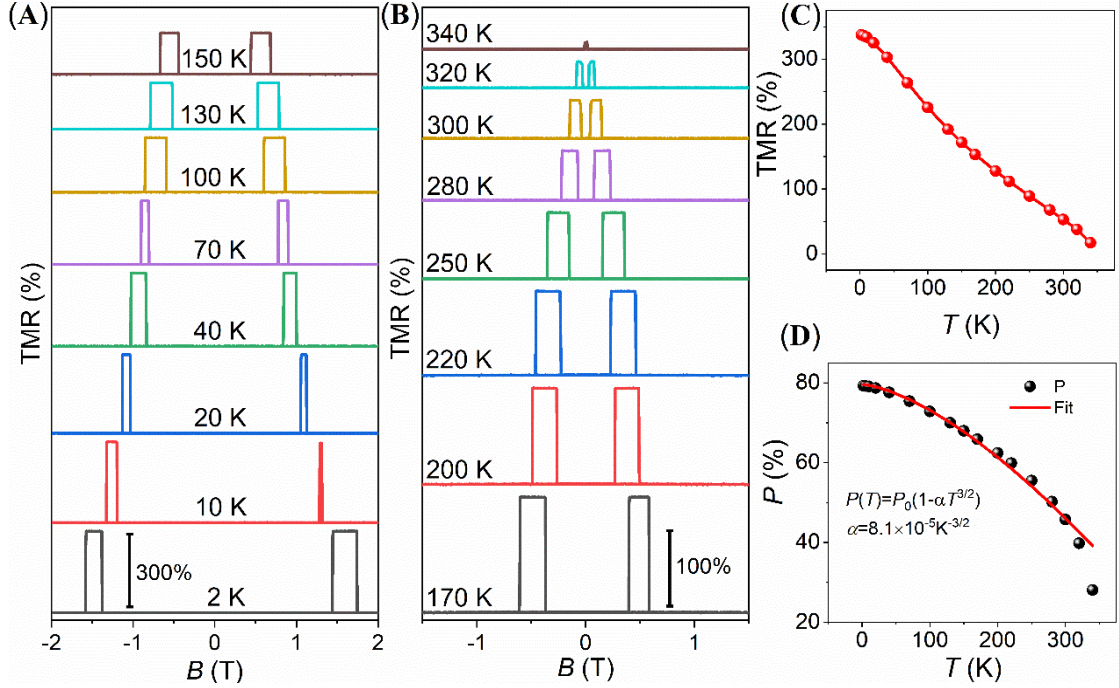

Figure 3. Temperature evolution of TMR in the FGT/WSe$_2$/FGT MTJ. (A) The offset plot of TMR curves measured at various temperatures ranging from 2 K to 150 K. (B) The TMR curves of various temperatures ranging from 170 K to 340 K. All the TMR are measured with a 10 nA AC current. (C) The extracted TMR ratio as a function of temperature. (D) Temperature dependence of the spin polarization $P$ extracted from the TMR (black dots). The red curve line is the fit $P$ from the formula $P(T)=P_0(1-\alpha T^{3/2})$.

## Conclusions

In conclusion, the room-temperature tunable TMR is realized by constructing a high-quality interface in full vdW FGT/WSe$_2$/FGT MTJ. The prominent TMR can be detected by a current lower than 1 nA and the large spin polarization can be maintained both at low temperature and room temperature. The large TMR at low and room-temperature implies the great potential of 2D magnetic vdW MTJ for spintronic applications. Importantly, room-temperature tunable TMR via DC bias provides a new perspective on realizing room-temperature vdW functional spintronic devices.

## Methods

Sample and device fabrication. FGT/WSe$_2$/FGT vdW heterostructures were fabricated by the dry transfer method in a glove box with $O_2$ and $H_2O$ concentrations below 1 ppm. All the top and bottom $Fe_3GaTe_2$ flakes, and the few layers of $WSe_2$ flakes were mechanically exfoliated from their bulk crystals using Scotch tape onto the polydimethylsiloxane. All the $Fe_3GaTe_2$ and $WSe_2$ crystals were grown by the flux method. The Cr/Au (2 nm/18 nm) electrodes were pre-fabricated on $SiO_2$/Si substrates using the standard lithography and the lift-off process. The bottom layer of $Fe_3GaTe_2$ flakes were first transferred to the $SiO_2$/Si substrates with Au electrodes through the transfer station, and part of $Fe_3GaTe_2$ was connected to the Au electrodes to serve as the bottom ferromagnetic electrode. Then the tunnel barrier of a few-layer $WSe_2$ flake was transferred onto the target bottom $Fe_3GaTe_2$. Finally, another $Fe_3GaTe_2$ flake was transferred onto the $WSe_2$ and also connected to another Au electrode to serve as the top ferromagnetic electrode. To isolate the MTJ from the atmosphere, the whole heterostructure area was encapsulated by h-BN flakes.

Spin valve measurement. All the electrical measurements were performed in a physical property measurement system (PPMS, Quantum Design). The magnetoresistance measurement of MTJ was performed with standard low-frequency lock-in techniques (Stanford Research Systems, SR830; Zurich Instruments, MFLI). The DC measurement of *I-V* curves was applied through the Keithley 2636 source meter.

Computational details. QUANTUM Espresso codes with plane wave basis set have been employed to calculate ground state electronic properties using spin resolved density functional theory (DFT).[34-35] We used generalized gradient approximation (GGA)[36] for Perdew-Burke-Ernzerhof (PBE)[37] exchange-correlation functional for accurate electron-ion interactions incorporating projector augmented wave pseudopotentials[38]. The unit cell consists of single layer of $WSe_2$ which is sandwiched between two layers of $Fe_3GaTe_2$. The atomic relaxation calculations are performed with 'DFT-D' van der Waal (vdW) correction method[39] and the optimized interlayer distance is found to be 3.4 Å. The convergence criteria for self-consistent calculations are achieved as convergence threshold becomes $9.6 \times 10^{-9}$ eV in company with force on each atom becomes $1.0 \times 10^{-4}$ eV/Å. Here, K point grid has been considered using Monk horst and Pack scheme within irreducible Brillouin zone.[40] Accordingly, in the first Brillouin zone 6×6×1 K point mesh for self-consistent calculations were used with an energy cut-off of 60 Ry. In addition, for non-self-consistent calculations, denser K point mesh as 18×18×1 is used. However, periodic image interactions have been taken care of by adding vacuum of 30.3 Å along Z direction.

## Supporting Information

Additional supporting information may be found in the online version of the article at the publisher's website.

## Acknowledgments

The research was supported by the Competitive Research Program of Singapore National Research Foundation (CRP Awards No. NRF-CRP22-2019-0004, No. NRF-CRP23-2019-0002). Haiyang Pan thanks the financial support from the National Natural Science Foundation of China (Grant No. 12104391) and the program of the China Scholarships Council (File No. 202008440015). Anil Kumar Singh and Pritam

Deb thank the financial support from ASEAN Collaborative project (Grant No. SERB/F/2909/2021-2022).

## Conflict of Interest

The authors declare no conflict of interest.